\documentclass[10pt]{iopart}

\usepackage{graphicx}
\usepackage{dcolumn}
\usepackage{bm}
\usepackage{color}

\begin{document}

\title[JT-60SA]{MHD stability of JT-60SA operation scenarios driven by passing energetic particles for a hot Maxwellian model}


\author{J. Varela}
\ead{jacobo.varela@nifs.ac.jp}
\address{National Institute for Fusion Science, National Institute of Natural Science, Toki, 509-5292, Japan}
\address{Universidad Carlos III de Madrid, 28911 Leganes, Madrid, Spain}
\author{K. Y. Watanabe}
\address{National Institute for Fusion Science, National Institute of Natural Science, Toki, 509-5292, Japan}
\author{K. Shinohara}
\address{National Institutes for Quantum and Radiological Science and Technology, Naka, Ibaraki 311-0193, Japan}
\author{M. Honda}
\address{National Institutes for Quantum and Radiological Science and Technology, Naka, Ibaraki 311-0193, Japan}
\author{Y. Suzuki}
\address{National Institute for Fusion Science, National Institute of Natural Science, Toki, 509-5292, Japan}
\author{J. Shiraishi}
\address{National Institutes for Quantum and Radiological Science and Technology, Naka, Ibaraki 311-0193, Japan}
\author{D. A. Spong}
\address{Oak Ridge National Laboratory, Oak Ridge, Tennessee 37831-8071, USA}
\author{L. Garcia}
\address{Universidad Carlos III de Madrid, 28911 Leganes, Madrid, Spain}

\date{\today}

\begin{abstract}
We analyze the effects of the passing energetic particles on the resistive ballooning modes (RBM) and the energetic particle driven modes in JT-60SA plasma, which leads to the prediction of  the stability in N-NBI heated plasma. The analysis is performed using the code FAR3d that solves the reduced MHD equations describing the linear evolution of the poloidal flux and the toroidal component of the vorticity in a full 3D system, coupled with equations of density and parallel velocity moments for the energetic particle (EP) species assuming an averaged Maxwellian EP distribution fitted to the slowing down distribution, including the effect of the acoustic modes. The simulations show the possible destabilization of a $3/2-4/2$ TAE with a frequency ($f$) of $115$ kHz, a $6/4-7/4$ TAE with $f=98$ kHz and a $6/4$ or $7/4$ BAE with $f=57$ kHz in the ITER-like inductive scenario. If the energetic particle $\beta$ increases, Beta induced AEs (BAE), Toroidal AEs (TAE) and Elliptical AEs (EAE) are destabilized between the inner-middle plasma region, leading to the overlapping of AE of different toroidal families. If these instabilities coexist in the non-linear saturation phase the EP transport could be enhanced leading to a lower heating efficiency. For a hypothetical configuration based on the ITER-like inductive scenario but an center peaked EP profile, the EP $\beta$ threshold increases and several BAEs are destabilized in the inner plasma region, indicating an improved AE stability with respect to the off-axis peaked EP profile. In addition, the analysis of a hypothetical JT-60SA scenario with a resonant $q=1$ in the inner plasma region shows the destabilization of fishbones-like instabilities by the off-axis peaked EP profile. Also, the EPs have an stabilizing effect on the RBM, stronger as the population of EP with low energies (below $250$ keV) increases at the plasma pedestal.
\end{abstract}

%
%
%
%
%


\vspace{2pc}

\maketitle

\ioptwocol

\section{Introduction \label{sec:introduction}}

JT-60SA is a key milestone in the goal of commercial nuclear fusion energy \cite{1,2,3}, anticipating scientific and engineering challenges of ITER \cite{4} and DEMO \cite{5,6} devices. In JT-60SA plasma several ITER and DEMO-like operation scenarios will be tested \cite{7,8}: inductive (standard H-mode) scenarios similar to ITER baseline scenarios, advanced inductive scenarios (high $\beta$ and low magnetic shear) similar to ITER hybrid scenarios and steady state scenarios mimicking the plasma conditions of a nuclear fusion power plant \cite{9,10,11,12}. The present study will be dedicated to analysis of an ITER-like inductive scenario \cite{13}.

JT-60SA is a Tokamak with a major radius of $2.96$ m and a minor radius of $1.18$ m. The plasma current is $5.5$ MA and the on-axis magnetic field magnitude is $2.25$ T. The plasma volume is $132$ m$^3$ and the inductive pulse lasts around $100$ s. The plasma will be heated by up to $34$ MW of neutral beam (NB) and $7$ MW of electron cyclotron resonance heating (ECRH). Between the NBs, $12$ positive-ion-based NB (P-NBI) and two negative-ion-based NB (N-NBI). The N-NBI will inject $10$ MW Deuterium beams generating energetic particles (EP) with an energy of $500$ keV deposited in the middle plasma region and the magnetic axis \cite{14,15}.

The EP can generate instabilities enhancing the transport of fusion produced alpha particles, energetic neutral beams and ion cyclotron resonance heated particles (ICRF) \cite{16,17,18}, reducing the heating efficiency in tokamaks such as JET and DIII-D \cite{19,20,21,22}. These instabilities could be also triggered in JT-60SA plasma if there is a resonance between the unstable mode frequency and the EP drift, bounce or transit frequencies \cite{23}, as was observed in JT-60U \cite{24,25,26,27,28}. 

Alfv\' en Eigenmodes (AE) are driven in the spectral gaps of the shear Alfv\' en continua \cite{29,30}. The Alfv\' en eigenmode belong to different families ($n$ is the toroidal mode and $m$ the poloidal mode) linked to the frequency gaps produced by periodic variations of the Alfv\' en speed, for example: toroidicity induced Alfv\' en Eigenmodes (TAE) couple $m$ with $m+1$ modes \cite{31,32,33}, beta induced Alfv\' en Eigenmodes driven by compressibility effects (BAE) \cite{34}, Reversed-shear Alfv\' en Eigenmodes (RSAE) due to local maxima/minima in the safety factor $q$ profile \cite{35}, Global Alfv\' en Eigenmodes (GAE) observed in the minimum of the Alfv\' en continua \cite{36}, ellipticity induced Alfv\' en Eigenmodes (EAE) coupling $m$ with $m+2$ modes \cite{37,38} and noncircularity induced Alfv\' en Eigenmodes (NAE) coupling $m$ with $m+3$ or higher \cite{39,40}. In addition, energetic particle modes (EPM) can be unstable for frequencies in the shear Alfven continua if the continuum damping is not strong enough to stabilize them \cite{41,42,43,44,45}. Examples of EPM are the energetic-ion-driven resistive interchange mode (EIC) \cite{46,47,48} or the fishbone oscillations \cite{49}.

Another dangerous instability for JT-60SA performance is the resistive ballooning mode (RBM) \cite{50,51,52}, the driver of the type III edge localized modes (ELMs) \cite{53,54}. This kind of instability must be avoided to reduce the heat load on the divertor and plasma facing components.

We investigate the effects of the passing energetic particles on the RBM and EPM/AE in JT-60SA plasma, which could lead to the prediction of the stability in on- and off-axis N-NBI heated plasmas. To that end, a set of simulations are performed using the FAR3d code \cite{55,56,57} analyzing ITER-like inductive scenarios. In addition, a JT60SA scenario with a resonant $q=1$ at the inner plasma region is analyzed, studying the stability of fish-bones. FAR3d code variables evolve starting from an equilibria calculated by the VMEC code \cite{58}. The numerical model solves the reduced linear resistive MHD equations and the moment equations of the energetic ion density and parallel velocity \cite{59,60} including the linear wave-particle resonance effects required for Landau damping/growth. Since the numerical model is only applicable to growing or damped modes, these resonances are in the complex plane and incorporated by closure relations that have been derived by performing contour integral deformations around singularities as, for example, used in evaluations of the plasma dispersion function. The parallel momentum response of the thermal plasma is included, as  required for coupling to the geodesic acoustic waves \cite{61}. A Maxwellian equilibrium EP distribution is chosen which has the same second moment, effective EP temperature, as the slowing down distribution as defined in the appendix.

Not all the resonances identified by the simulations should lead to the destabilization of AEs in the experiment, because the drive is determined by the gradient of the phase space distribution and the gradient depends on the phase space shape of the distribution function of the EP. Nevertheless, this information is useful for future optimization studies. It is important to point out that the model reproduces the destabilizing effect of the passing EP, hence the effect of highly anisotropic beams or ICRF driven EP cannot be modeled by the present version of the code. However, the pitch angle of the EP generated by the tangential N-NBI in JT60SA plasma should be small. In addition, due to the Maxwellian distribution function used for the EP model, the co-EP (pitch angle is $0$) and ctr-EP (pitch angle is $\pi$ radians) lead to the same resonance (same frequency and growth rate) although the mode propagates in the opposite direction. Thus, the observed modes can be caused by co-passing EPs generated by the N-NBI.

This paper is organized as follows. The model equations, numerical scheme and equilibrium properties are described in section \ref{sec:model}. The stability analysis of RBM and AEs in a ITER-like inductive scenario and an hypothetical on-axis EP deposition case is done in section \ref{sec:CaseA}. Next, the stability of the fish-bones is studied in a hypothetical scenario where the safety factor is below unity at the magnetic axis in section \ref{sec:CaseB}. Finally, the conclusions of this paper are presented in section \ref{sec:conclusions}.

\section{Equations and numerical scheme \label{sec:model}}

Following the method employed in Ref.\cite{62}, a reduced set of equations for high-aspect ratio configurations and moderate $\beta$-values (of the order of the inverse aspect ratio) is derived retaining the toroidal angle variation, based upon an exact three-dimensional equilibrium that assumes closed nested flux surfaces. The effect of the energetic particle population in the plasma stability is included through moments of the fast ion kinetic equation truncated with a closure relation \cite{63}, describing the evolution of the energetic particle density ($n_{f}$) and velocity moments parallel to the magnetic field lines ($v_{||f}$). The coefficients of the closure relation are selected to match analytic TAE growth rates based upon a two-pole approximation of the plasma dispersion function.

The model formulation assumes high aspect ratio, medium $\beta$ (of the order of the inverse aspect ratio $\varepsilon=a/R_0$), small variation of the fields and small resistivity. The plasma velocity and perturbation of the magnetic field are defined as
\begin{equation}
 \mathbf{v} = \sqrt{g} R_0 \nabla \zeta \times \nabla \Phi, \quad\quad\quad  \mathbf{B} = R_0 \nabla \zeta \times \nabla \tilde\psi,
\end{equation}
where $\zeta$ is the toroidal angle, $\Phi$ is a stream function proportional to the electrostatic potential, and $\tilde \psi$ is the perturbation of the poloidal flux.

The equations, in dimensionless form, are
\begin{equation}
\frac{\partial \tilde \psi}{\partial t} =  \sqrt{g} B \nabla_\| \Phi  + \frac{\eta}{S} \tilde J^\zeta
\end{equation}
\begin{eqnarray} 
\frac{{\partial \tilde U}}{{\partial t}} =  - v_{\zeta,eq} \frac{\partial \tilde U}{\partial \zeta} \nonumber\\
+ \sqrt{g} B  \nabla_{||} \tilde J^{\zeta} - \frac{1}{\rho} \left( \frac{\partial J_{eq}}{\partial \rho} \frac{\partial \tilde \psi}{\partial \theta} - \frac{\partial J_{eq}}{\partial \theta} \frac{\partial \tilde \psi}{\partial \rho}    \right) \nonumber\\
- {\frac{\beta_0}{2\varepsilon^2} \sqrt{g} \left( \nabla \sqrt{g} \times \nabla \tilde p \right)^\zeta } -  {\frac{\beta_f}{2\varepsilon^2} \sqrt{g} \left( \nabla \sqrt{g} \times \nabla \tilde n_f \right)^\zeta }
\end{eqnarray} 
\begin{eqnarray}
\label{pressure}
\frac{\partial \tilde p}{\partial t} = - v_{\zeta,eq} \frac{\partial \tilde p}{\partial \zeta} + \frac{dp_{eq}}{d\rho}\frac{1}{\rho}\frac{\partial \tilde \Phi}{\partial \theta} \nonumber\\
 +  \Gamma p_{eq}  \left[{\left( \nabla \sqrt{g} \times \nabla \tilde \Phi \right)^\zeta - \nabla_\|  \tilde v_{\| th} }\right] 
\end{eqnarray} 
\begin{eqnarray}
\label{velthermal}
\frac{{\partial \tilde v_{\| th}}}{{\partial t}} = - v_{\zeta,eq} \frac{\partial \tilde v_{||th}}{\partial \zeta} -  \frac{\beta_0}{2n_{0,th}} \nabla_\| p 
\end{eqnarray}
\begin{eqnarray}
\label{nfast}
\frac{{\partial \tilde n_f}}{{\partial t}} = - v_{\zeta,eq} \frac{\partial \tilde n_{f}}{\partial \zeta} - \frac{v_{th,f}^2}{\varepsilon^2 \omega_{cy}}\ \Omega_d (\tilde n_f) - n_{f0} \nabla_\| \tilde v_{\| f}   \nonumber\\
-  n_{f0} \, \Omega_d (\tilde \Phi) + n_{f0} \, \Omega_* (\tilde  \Phi)  
\end{eqnarray}
\begin{eqnarray}
\label{vfast}
\frac{{\partial \tilde v_{\| f}}}{{\partial t}} = - v_{\zeta,eq} \frac{\partial \tilde v_{||f}}{\partial \zeta}  -  \frac{v_{th,f}^2}{\varepsilon^2 \omega_{cy}} \, \Omega_d (\tilde v_{\| f}) \nonumber\\
- \left( \frac{\pi}{2} \right)^{1/2} v_{th,f} \left| \nabla_\| \tilde v_{\| f}  \right| - \frac{v_{th,f}^2}{n_{f0}} \nabla_\| n_f + v_{th,f}^2 \, \Omega_* (\tilde \psi) 
\end{eqnarray}
Equation (2) is derived from Ohm$'$s law coupled with Faraday$'$s law, equation (3) is obtained from the toroidal component of the momentum balance equation after applying the operator $\vec{\nabla} \times \sqrt{g}\vec{X}$ (with $\vec{X}$ a given vectorial variable), equation (4) is obtained from the thermal plasma continuity equation with compressibility effects and equation (5) is obtained from the parallel component of the momentum balance. Here, $U =  \sqrt g \left[{ \nabla  \times \left( {\rho _m \sqrt g {\bf{v}}} \right) }\right]^\zeta$ is the toroidal component of the vorticity, $\rho_m$ the ion and electron mass density, $\rho = \sqrt{\phi_{N}}$ the effective radius with $\phi_{N}$ the normalized toroidal flux and $\theta$ the poloidal angle. The perturbation of the toroidal current density $\tilde J^{\zeta}$ is defined as:
\begin{eqnarray}
\tilde J^{\zeta} =  \frac{1}{\rho}\frac{\partial}{\partial \rho} \left(-\frac{g_{\rho\theta}}{\sqrt{g}}\frac{\partial \tilde \psi}{\partial \theta} + \rho \frac{g_{\theta\theta}}{\sqrt{g}}\frac{\partial \tilde \psi}{\partial \rho} \right) \nonumber\\
- \frac{1}{\rho} \frac{\partial}{\partial \theta} \left( -\frac{g_{\rho\rho}}{\sqrt{g}}\frac{1}{\rho}\frac{\partial \tilde \psi}{\partial \theta} + \rho \frac{g_{\rho \theta}}{\sqrt{g}}\frac{\partial \tilde \psi}{\partial \rho} \right)
\end{eqnarray}
$v_{||th}$ is the parallel velocity of the thermal particles and $v_{\zeta,eq}$ is the equilibrium toroidal rotation. $n_{f}$ is normalized to the density at the magnetic axis $n_{f_{0}}$, $\Phi$ to $a^2B_{0}/\tau_{A0}$ and $\tilde\Psi$ to $a^2B_{0}$ with $\tau_{A0}$ the Alfv\' en time $\tau_{A0} = R_0 (\mu_0 \rho_m)^{1/2} / B_0$. The radius $\rho$ is normalized to plasma minor radius $a$; the resistivity to $\eta_0$ (its value at the magnetic axis); the time to the Alfv\' en time; the magnetic field to $B_0$ (the averaged value at the magnetic axis); and the pressure to its equilibrium value at the magnetic axis. The Lundquist number $S$ is the ratio of the resistive time $\tau_{R} = a^2 \mu_{0} / \eta_{0}$ to the Alfv\' en time. $\rlap{-} \iota$ is the rotational transform, $v_{th,f} = \sqrt{T_{f}/m_{f}}$ the energetic particle thermal velocity normalized to the Alfv\' en velocity in the magnetic axis $v_{A0}$ and $\omega_{cy}$ the energetic particle cyclotron frequency normalized to $\tau_{A0}$. $q_{f}$ is the charge, $T_{f}$ the temperature and $m_{f}$ the mass of the energetic particles. The $\Omega$ operators are defined as:
\begin{eqnarray}
\label{eq:omedrift}
\Omega_d = \frac{1}{2 B^4 \sqrt{g}}  \Bigg\{ \left( \frac{I}{\rho} \frac{\partial B^2}{\partial \zeta} - J \frac{1}{\rho} \frac{\partial B^2}{\partial \theta} \right) \frac{\partial}{\partial \rho} \nonumber\\
-  \left( \rho \beta_* \frac{\partial B^2}{\partial \zeta} - J \frac{\partial B^2}{\partial \rho} \right) \frac{1}{\rho} \frac{\partial}{\partial \theta} \nonumber\\ 
+ \left( \rho \beta_* \frac{1}{\rho} \frac{\partial B^2}{\partial \theta} -  \frac{I}{\rho} \frac{\partial B^2}{\partial \rho} \right) \frac{\partial}{\partial \zeta} \Bigg\}
\end{eqnarray}

\begin{eqnarray}
\label{eq:omestar}
\Omega_* = \frac{1}{B^2 \sqrt{g}} \frac{1}{n_{f0}} \frac{d n_{f0}}{d \rho} \left( \frac{I}{\rho} \frac{\partial}{\partial \zeta} - J \frac{1}{\rho} \frac{\partial}{\partial \theta} \right) 
\end{eqnarray}
Here the $\Omega_{d}$ operator is constructed to model the average drift velocity of a passing particle ($v_{\perp} = 0$) and $\Omega_{*}$ models its diamagnetic drift frequency. We also define the parallel gradient and curvature operators as
\begin{equation}
\label{eq:gradpar}
\nabla_\| f = \frac{1}{B \sqrt{g}} \left( \frac{\partial \tilde f}{\partial \zeta} - \rlap{-} \iota \frac{\partial \tilde f}{\partial \theta} - \frac{\partial f_{eq}}{\partial \rho}  \frac{1}{\rho} \frac{\partial \tilde \psi}{\partial \theta} + \frac{1}{\rho} \frac{\partial f_{eq}}{\partial \theta} \frac{\partial \tilde \psi}{\partial \rho} \right)
\end{equation}
\begin{equation}
\label{eq:curv}
\sqrt{g} \left( \nabla \sqrt{g} \times \nabla \tilde f \right)^\zeta = \frac{\partial \sqrt{g} }{\partial \rho}  \frac{1}{\rho} \frac{\partial \tilde f}{\partial \theta} - \frac{1}{\rho} \frac{\partial \sqrt{g} }{\partial \theta} \frac{\partial \tilde f}{\partial \rho}
\end{equation}
with the Jacobian of the transformation,
\begin{equation}
\label{eq:Jac}
\frac{1}{\sqrt{g}} = \frac{B^2}{(J- \rlap{-} \iota I)}
\end{equation}

Equations~\ref{pressure} and~\ref{velthermal} introduce the parallel momentum response of the thermal plasma. These are required for coupling to the geodesic acoustic waves, accounting for the geodesic compressibility in the frequency range of the geodesic acoustic mode (GAM) \cite{64,65}.

Equilibrium flux coordinates $(\rho, \theta, \zeta)$ are used. Here, $\rho$ is a generalized radial coordinate proportional to the square root of the toroidal flux function, and normalized to the unity at the edge. The flux coordinates used in the code are those described by Boozer \cite{66}, and $\sqrt g$ is the Jacobian of the coordinate transformation. All functions have equilibrium and perturbation components represented as: $ A = A_{eq} + \tilde{A} $.

The FAR3D code uses finite differences in the radial direction and Fourier expansions in the two angular variables. In the present study, the numerical scheme used to resolve the linear equations is an eigen-value solver that can calculate the stable and unstable modes (sub-dominant modes).

The representation of the eigenfunctions ($f$) in FAR3d code is done internally using cosine and sine components, but for the plots used in this paper we express these in terms of real (R) and imaginary (I) components:
\begin{eqnarray} 
Re[f(\rho,\theta, \zeta, t)] = Re[\sum_{m,n} (f^{R}_{mn}(\rho)+i f^{I}_{mn}(\rho)) \nonumber\\
(cos\omega_{R}t - isin\omega_{R}t)e^{i(m\theta + n\zeta)}]
\end{eqnarray}

The present model was already used to study the stability of RBM in DIII-D discharges with high poloidal $\beta$ \cite{67}, AE stability in DIII-D, ITER, LHD and TJ-II \cite{68,69,70,71} as well as the EIC in LHD \cite{48}, indicating reasonable agreement with the observations.

\subsection{Equilibrium properties}

Two fixed boundary results from the VMEC equilibrium code \cite{58}, reproducing an ITER-like inductive scenario (cases A) and a hypothetical scenario with a resonant $q=1$ at the inner plasma region (case B). Table~\ref{Table:1} shows the main parameters of the Deuterium thermal plasma and the details of the EP injected by the Deuterium N-NBI of the cases A and B. The magnetic field at the magnetic axis is $2.25$ T and the averaged inverse aspect ratio is $\varepsilon=0.4$.

\begin{table*}[t]
\centering
\begin{tabular}{c c c c c c c c c}
\hline
Case & $T_{th,e}$ & $n_{th,i}$ & $\beta_{tot}$ & $v_{A0}$ & $I_{p}$ & $T_{f,max}$ & $n_{f,max}$ & $\beta_{f,max}$ \\
& (keV) & ($10^{20}$ m$^{-3}$) & ($\%$) & ($10^{6}$ m/s) & MA & (keV) &  ($10^{20}$ m$^{-3}$) & ($\%$)\\ \hline
A & 8.3 & 0.77 & 9.5 & 3.96 & 4.6 & 280 & 0.0029 & 0.5\\
B & 5.9 & 0.60 & 9.1 & 4.48 & 3.5 & 280 & 0.0027 & 0.5\\ \hline
\end{tabular}
\caption{Thermal plasma and EP parameters of the models. The first column indicates the model name, the second column the electron temperature, the third column the ion density, the fourth column is the total $\beta$, the fifth column the Alfv\' en velocity, the sixth column the plasma current, the seventh column the EP temperature, the eight column the EP density and the ninth column the EP $\beta$.} \label{Table:1}
\end{table*}

The energy of the injected particles by the tangential N-NBI is $500$ keV, but we take the nominal energy $T_{f} = 280$ keV ($v_{th,f} = 3.66 \cdot 10^{6}$ m/s) resulting in an averaged Maxwellian energy equal to the average energy of a slowing-down distribution (see the Appendix for further information). It should be noted that a model with an averaged EP Maxwellian distribution and a single EP energy cannot reproduce the same resonances triggered by a slowing down EP distribution as mentioned in the introduction.

\begin{figure}[h!]
\centering
\includegraphics[width=0.45\textwidth]{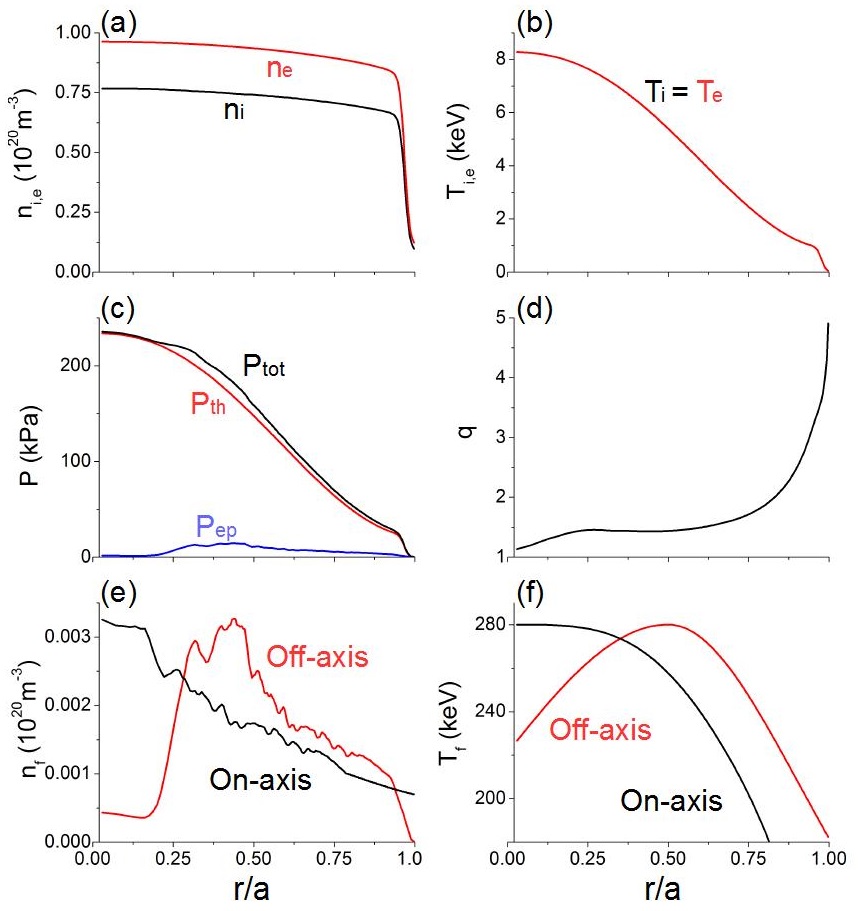}
\caption{Case A. Thermal plasma profiles: (a) Electron and ion density (a), electron and ion temperature (b), total, thermal and EP pressure (c), safety factor (d). EP profiles if the N-NBI is deposited on-axis (black line) or off-axis (red line): density (e) and EP energy (f).}\label{FIG:1}
\end{figure}

\begin{figure}[h!]
\centering
\includegraphics[width=0.45\textwidth]{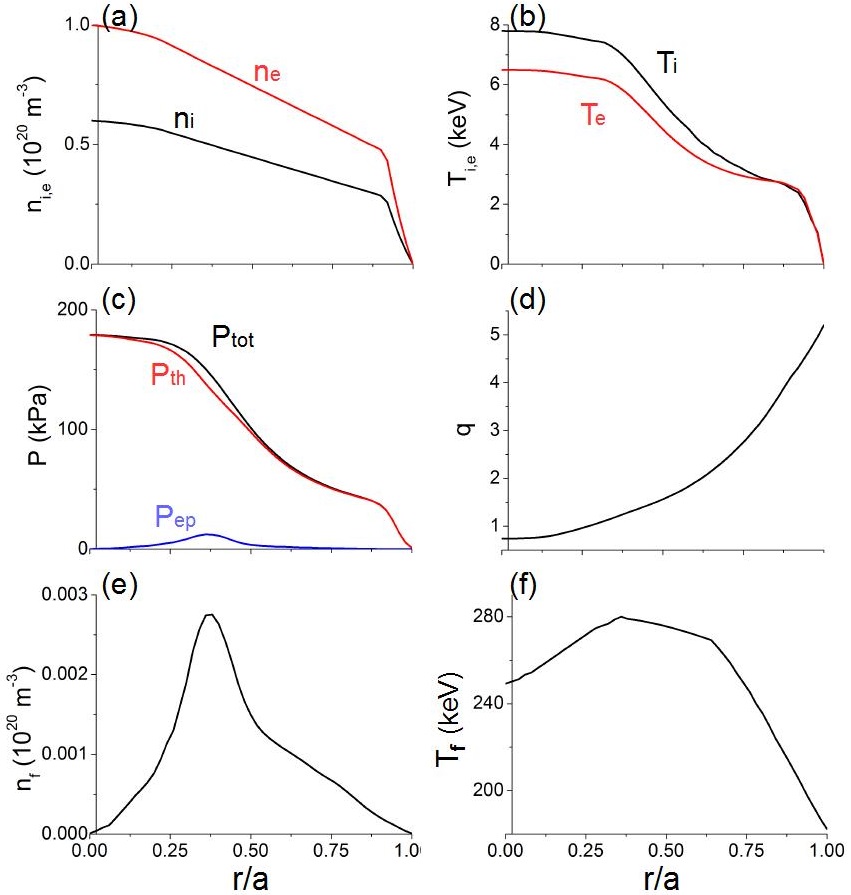}
\caption{Case B. Thermal plasma profiles: (a) Electron and ion density (a), electron and ion temperature (b), total, thermal and EP pressure (c), safety factor (d). EP profiles if the N-NBI is deposited off-axis: density (e) and nominal EP energy (f).}\label{FIG:2}
\end{figure}

The thermal plasma and EP profiles in the ITER-like inductive scenario and the hypothetical configuration with on-axis EP profile is shown in figure~\ref{FIG:1}. The thermal plasma and EP profiles of the scenario with a resonant $q=1$ in the inner plasma region is shown in figure~\ref{FIG:2}.

\begin{figure}[h!]
\centering
\includegraphics[width=0.45\textwidth]{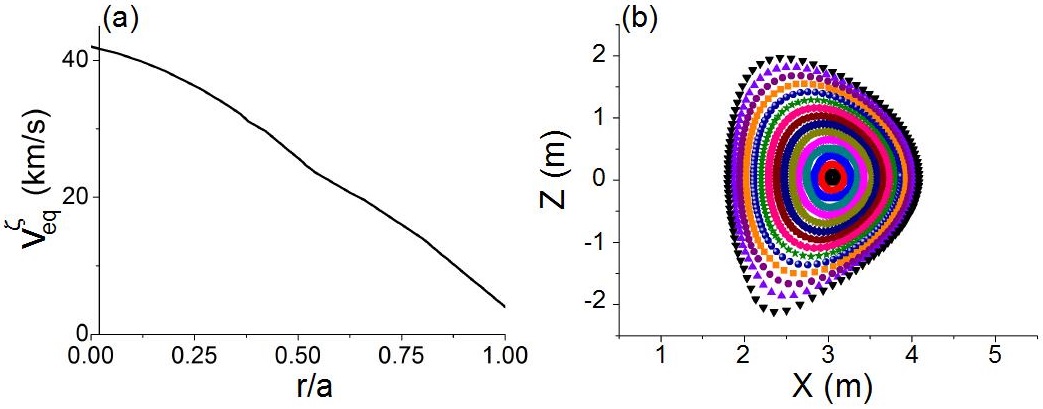}
\caption{Hypothetical equilibrium toroidal rotation profile (a) and the magnetic surfaces of the case A (b).}\label{FIG:3}
\end{figure}

The thermal plasma and EP profiles in the off-axis case A are consistent with the ITER-like inductive scenarios \cite{9}. The EP density profile is calculated from the expected EP pressure profile in ITER-like inductive scenarios (off-axis case A) although the EP energy profile is assumed based on the case of DIII-D and ITER obtained by TRANSP code \cite{72}. Regarding the on-axis case, both EP density and energy profiles are hypothetical, obtained displacing the local maxima of the EP density and energy to the magnetic axis. The profile of the EP nominal energy has the maxima at the radial location where the beam is injected and decreases smoothly away from the injection region. Figure~\ref{FIG:3} shows the hypothetical equilibrium toroidal rotation profile included in the model and the magnetic surfaces of the case A (similar for the case B, data not shown). The main difference between the ITER-like inductive scenario and the hypothetical case with a resonant $q=1$ rational surface is the q profile, thus the same operational regime of the NBI is assumed in both scenarios, reason why the EP profiles are similar.

\subsection{Simulations parameters}

The dynamic toroidal modes ($n$) in the simulations range from $n=1$ to $4$ and the dynamic poloidal modes ($m$) are selected to cover all the resonant rational surfaces. The dynamic mode selection changes between the case A and B because the safety factor profiles are different. Table~\ref{Table:2} shows the equilibrium and dynamic modes for the cases A and B.

\begin{table}[t]
\centering
\begin{tabular}{c c c}
\hline
n & m (case A) & m (case B) \\ \hline
0 & [0,11] & [0,11] \\
1 & [1,4] & [1,5] \\
2 & [3,10] & [1,10] \\
3 & [4,15] & [2,15] \\
4 & [6,20] & [3,20] \\ \hline
\end{tabular}
\caption{Equilibrium and dynamic modes in the simulations. The first column shows the toroidal modes, the second columns the poloidal modes in case A and the third column the poloidal modes in case B} \label{Table:2}
\end{table}

The closure of the kinetic moment equations (6) and (7) breaks the MHD parities so both parities must be included for all the dynamic variables. The convention of the code with respect to the Fourier decomposition is, in the case of the pressure eigenfunction, that $n>0$ corresponds to $cos(m\theta + n \zeta)$ and $n<0$ corresponds to $sin(-m\theta - n \zeta)$. For example, the Fourier component for mode $2/1$ is $\cos(2\theta + 1\zeta)$ and for the mode $-2/-1$ is $\sin(2\theta + 1\zeta)$. The magnetic Lundquist number is assumed $S=5\cdot 10^6$.

\section{ITER-like inductive scenario \label{sec:CaseA}}

This section is dedicated to study the stability of RBM, current gradient driven modes (CGDM) and AEs in an ITER-like inductive scenario of JT-60SA (off-axis case A).

\begin{figure}[h!]
\centering
\includegraphics[width=0.45\textwidth]{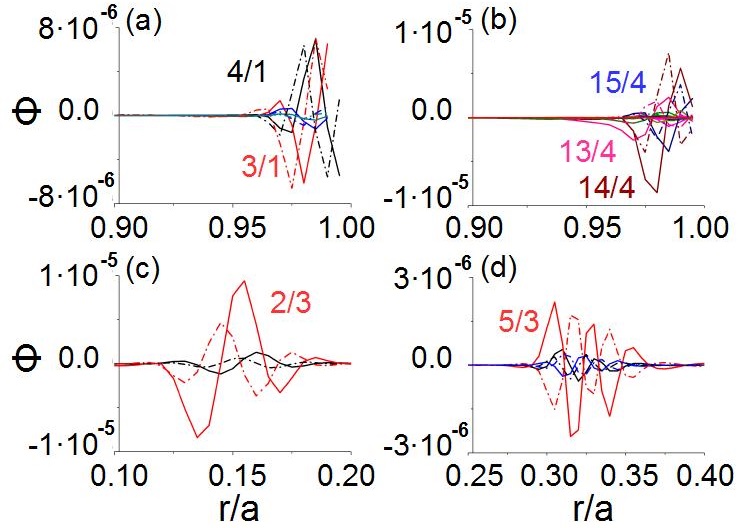}
\caption{Eigen-function of the (a) $n=1$ RBM, (b) $n=4$ RBM, (c) $n=2$ CGDM and (d) $n=3$ CGDM in the ITER-like inductive scenario. Solid lines indicate the real component and the dot-dashed lines the imaginary component of the eigenfunction.}\label{FIG:4}
\end{figure}

Figure~\ref{FIG:4}a and b show the RBM eigen-function of the $n=1$ and $4$ RBM for the ITER-like inductive scenario ($\beta_{f} = 0.01$). These modes are triggered by pressure gradients. They show symmetry with respect to mode parities and small real frequencies, as is characteristic for MHD ballooning modes. The eigen-function of the $n=1$ and $4$ RBM indicate instabilities located at the plasma pedestal ($r/a \approx 0.97$) destabilized by the $q=4$ rational surface with large poloidal couplings, typical structure of the RBM. It should be noted that these modes are stable if the magnetic Lundquist number of the simulation is $S=7\cdot 10^7$, similar to the experimental conditions (lower plasma resistivity). Previous numerical analysis indicated that only high n ballooning modes are unstable in ITER-like inductive scenario \cite{73} and the global (low n) external kink modes are marginally stable \cite{3,74}. Figure~\ref{FIG:4}c and d show the eigenfunction of the $n=2$ and $n=3$ CGDM, destabilized in the inner plasma region with a frequency of $8$ and $10$ kHz, respectively. If the magnetic Lundquist number of the simulation increases to $S=10^8$, similar to the experimental conditions, these modes are marginally stable.

\begin{figure}[h!]
\centering
\includegraphics[width=0.5\textwidth]{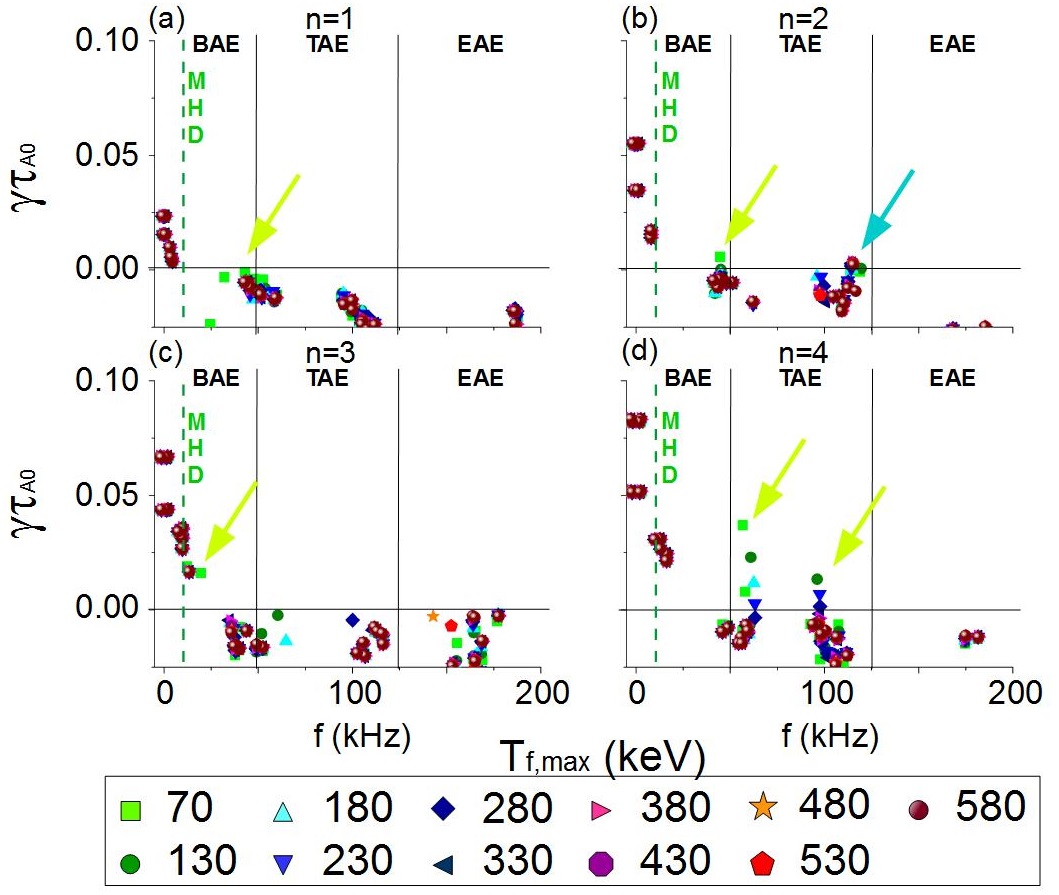}
\caption{Growth rate and frequency of the AEs for different EP energies in the ITER-like inductive scenario. The horizontal solid black line indicates the transition between stable (negative growth rate) and unstable (positive growth rate) modes. The vertical dashed green line indicates the range of frequencies of the AE (right side) and the pressure gradient driven modes (left side). The vertical solid black lines indicate the range of frequencies of the BAE, TAE and EAE. The yellow (cyan) arrows indicates the modes that shows an increase (decrease) of the growth rate as the EP energy decreases.}\label{FIG:5}
\end{figure}

Figure~\ref{FIG:5} shows the growth rate and frequency of the AEs for different EP energies (same profile as fig.~\ref{FIG:1}f although scaling the maxima of the EP energy , please see fig~\ref{FIG:14}b). A study of this type is required to analyze all the possible resonances between the EP and bulk plasma in the ITER-like inductive scenario. This parametric study is needed to generalize the analysis to other possible resonances based on our mode (passing EP with a Maxwellian distribution). It should be noted that the EP $\beta$ is fixed in the simulations, so the decrease/increase of the EP energy is compensated by an increase/decrease of the EP density. The simulations show a marginally stable n=1 BAE with $43$ kHz if $T_{f,max} < 70$ keV, an unstable $n=2$ BAE with $45$ kHz if $T_{f,max} < 130$ keV and a marginally unstable TAE with $115$ kHz if $T_{f,max} > 280$ keV, an unstable $n=3$ BAE with $20$ kHz if $T_{f,max} < 130$ keV as well as an unstable $n=4$ BAE with $57$ kHz if $T_{f,max} < 280$ keV and a TAE with $96$ kHz if $T_{f,max} < 330$ keV. The $n=4$ BAE as well as the $n=2$ and $n=4$ TAEs are destabilized by EP with an energy and density consistent with the EP population in the ITER-like inductive scenario, thus these instabilities could be triggered. On the other hand, the $n=1$ to $n=3$ BAEs are destabilized only if there is a large population of EP with $T_{f,max} < 130$ keV, although this should not be the case for the EP generated by the tangential N-NBI in JT60SA, thus these modes may be stable.

\begin{figure}[h!]
\centering
\includegraphics[width=0.5\textwidth]{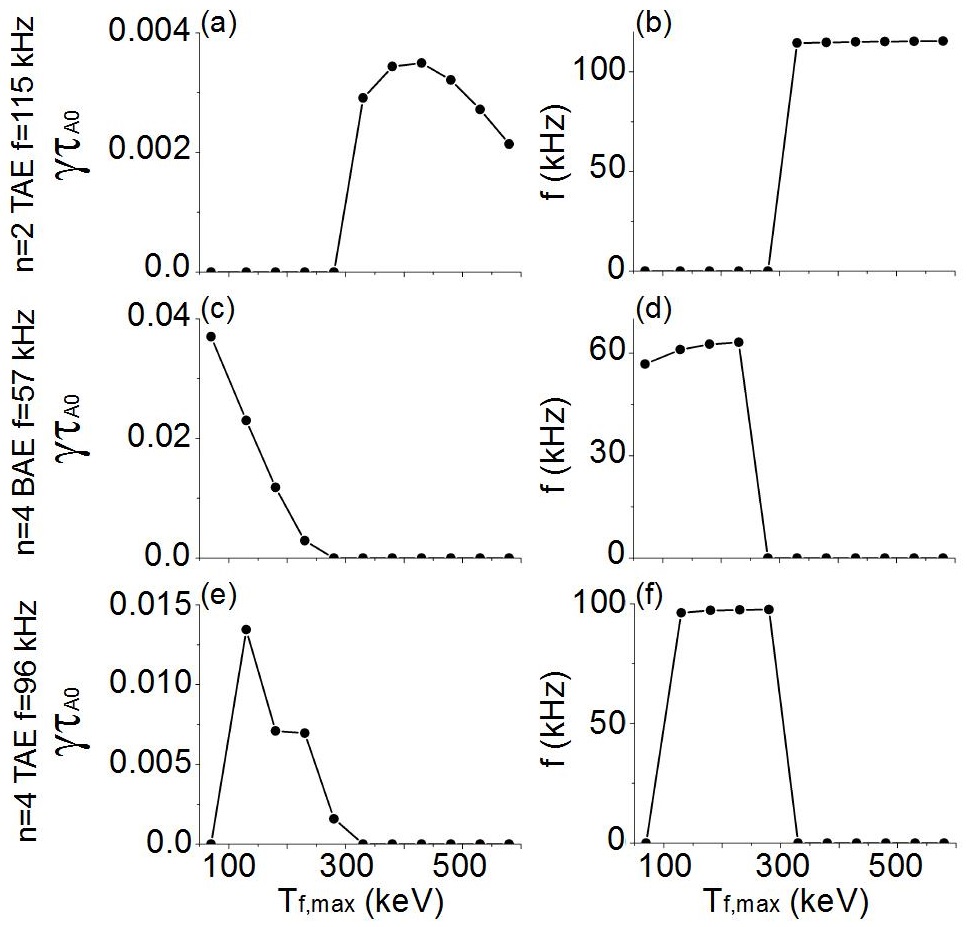}
\caption{Growth rate and frequency of the AEs for different EP energies in the ITER-like inductive scenario for the $n=2$ TAE (panels a and b), $n=4$ BAE with $f=57$ kHz (panels c and d) and $n=4$ TAE with $f=96$ kHz (panels e and f).}\label{FIG:6}
\end{figure}

To analyze in more detail the effect of the EP energy on the $n=4$ BAE, $n=2$ TAE and $n=4$ TAE stability, figure~\ref{FIG:6} indicates how the $n=2$ and $4$ AEs growth rate and frequency change as the EP energy is scanned from $70$ to $580$ keV. Once the TAEs are destabilized, the instability frequency remains almost constant although the growth rate changes, showing a different maximum of the growth rate with respect to the EP energy. The $n=4$ BAE and TAE can be destabilized by large populations of low energy EP (yellow arrows in fig.~\ref{FIG:5}d); this trend is associated with a decrease of the ratio $v_{th,f}/v_{A}$, that is to say, the EP thermalized velocity is far from the super-Alfvenic condition ($v_{th,f}/v_{A} > 1$) where the EP are weakly resonant or non-resonant \cite{68}. On the other hand, the $n=2$ TAE is unstable if the EP energy increases (cyan arrow in fig.~\ref{FIG:5}b), indicating that the high frequency AEs (above $100$ kHz) are easily destabilized if $v_{th,f}$ is closer to the super-Alfvenic condition even for a small population of EP.

\begin{figure}[h!]
\centering
\includegraphics[width=0.5\textwidth]{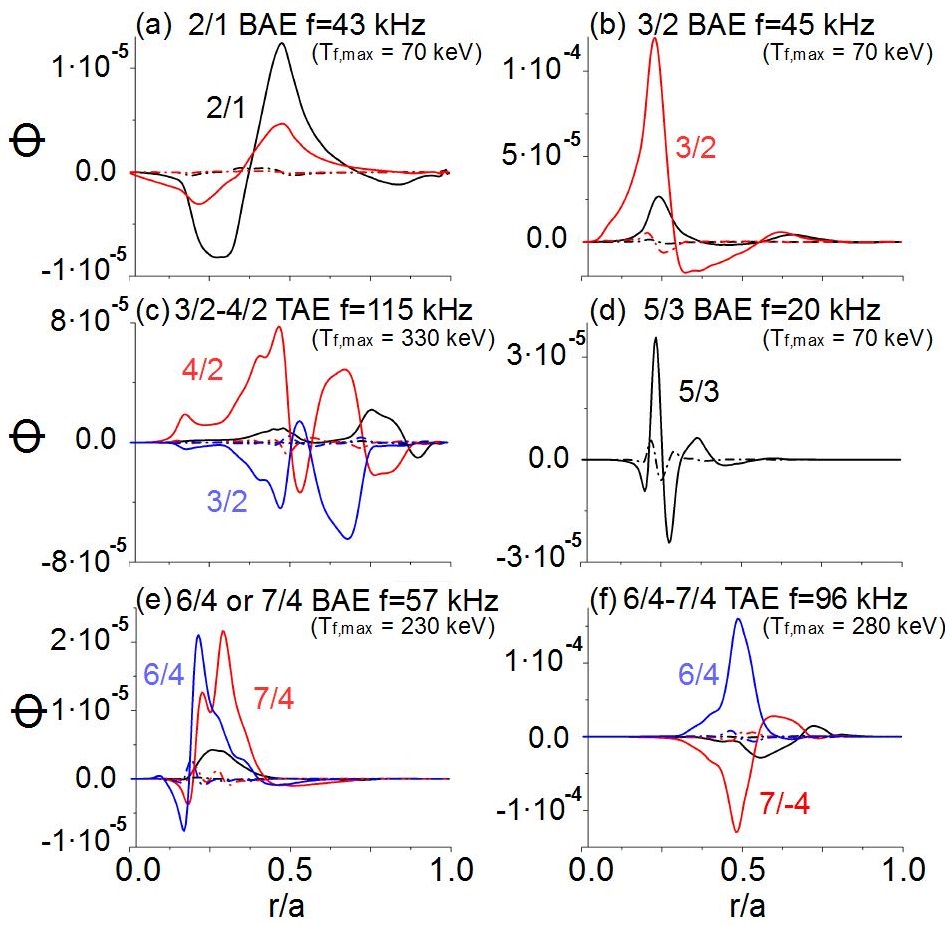}
\caption{Eigenfunction of the AEs unstable in the ITER-like inductive scenario for different EP energies: $2/1$ BAE (a) with $43$ kHz if $T_{f,max} = 70$ keV, $3/2$ BAE with $45$ kHz if $T_{f,max} = 70$ keV, $3/2-4/2$ TAE with $115$ kHz if $T_{f,max} = 330$ keV (c), $5/3$ BAE with $20$ kHz if $T_{f,max} = 70$ keV (d), $6/4-7/4$ TAE with $57$ kHz if $T_{f,max} = 230$ keV (e) and $6/4-7/4$ TAE with $96$ kHz if $T_{f,max} = 280$ keV (f). Solid lines indicate the real component and the dot-dashed lines the imaginary component of the eigenfunction.}\label{FIG:7}
\end{figure}

\begin{figure}[h!]
\centering
\includegraphics[width=0.45\textwidth]{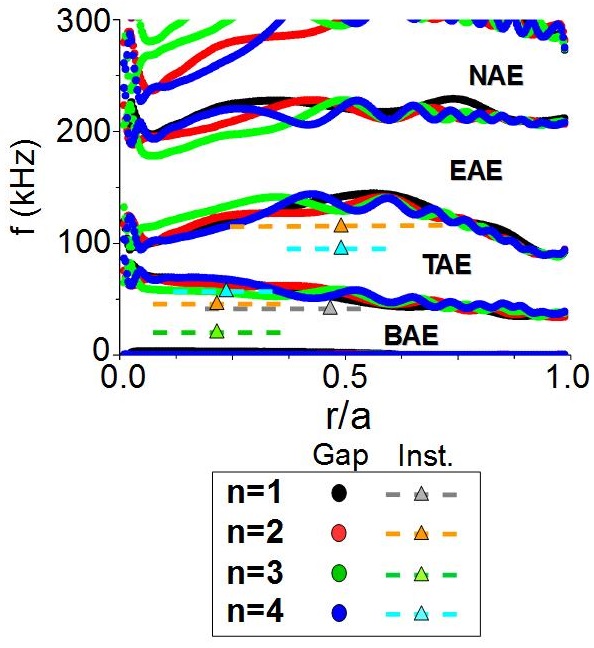}
\caption{Alfven gaps of the ITER-like inductive scenario. The dashed lines indicate the width of the unstable AEs and the triangles the eigen-function maxima.}\label{FIG:8}
\end{figure}

Figure~\ref{FIG:7} shows the eigenfunction of the AEs unstable in the ITER-like inductive scenario for different EP energies. There is a marginally stable $2/1$ BAE located in the inner-middle plasma region ($T_{f,max} = 70$ keV, panel a), an unstable $3/2$ BAE in the inner plasma region ($T_{f,max} = 70$ keV, panel b), a $3/2-4/2$ TAE in the middle plasma region ($T_{f,max} = 330$ keV, panel c), a $5/3$ BAE in the inner plasma region ($T_{f,max} = 70$ keV, panel d), $6/4-7/4$ TAEs with $64$ kHz ($T_{f,max} = 230$ keV, panel e) and $96$ kHz ($T_{f,max} = 280$ keV, panel f) are located in the inner and middle plasma region, respectively. Figure~\ref{FIG:8} indicates the Alfven gaps of the ITER-like inductive scenario calculated by the Stellgap code including the effect of the sound wave \cite{75}, adding the width of the AEs calculated by the FAR3d code at the frequency ranges where these instabilities are triggered. For this application of Stellgap the sound wave spectrum is simplified by using the 'slow sound' approximation \cite{76}. This approximations retains the BAE gap as shown, but suppresses most of the lower frequency BAAE gap structure. The simulations indicate that, if these AEs can coexist in the non-linear (saturation) phase, there is an overlapping of the $n=1$ to $4$ BAEs, particularly in the frequency range of the $50$ kHz between the inner-middle plasma region. Consequently, the EP transport can limit the performance of JT-60SA because the heating efficiency of the EP is reduced. Here, the EP transport enhancement by the AE overlapping is understood as the addition of the EP transport caused by the instabilities individually. Non linear simulations are required to study the overlapping of the resonant interaction.

In summary, the stability of the $n=1$ to $4$ RBM, CGDM and AE is analyzed in the ITER-like inductive scenario. The simulations indicate that the RBM are stable and the CGDM are marginal stable. Also, weakly thermalized EP ($T_{f,max} \ge 330$ keV) can destabilize $3/2-4/2$ TAE although EP during the slowing down process ($T_{f,max} \le 280$ keV) can destabilize $6/4-7/4$ TAE and $6/4$ or $7/4$ BAEs.

\subsection{Effect of the EP $\beta$ on the AE stability}

Now, the stability of the AEs in an ITER-like inductive scenario of JT-60SA is analyzed with respect the EP $\beta$ (case A off-axis). This analysis is performed using the EP energy profile shown in fig.~\ref{FIG:1}f and increasing the EP density.

\begin{figure*}[h!]
\centering
\includegraphics[width=0.95\textwidth]{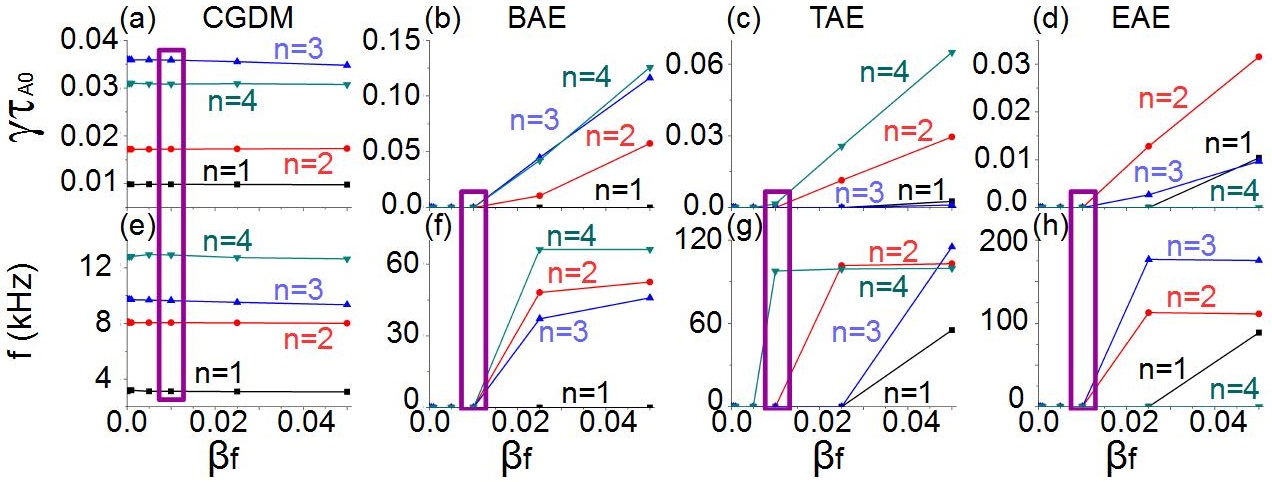}
\caption{Growth rate and frequency of the CGDM (a and e), BAE (b and f), TAE (c and g) and EAE (d and h) for different $\beta_{f}$. The purple box indicates the ITER-like inductive scenario.}\label{FIG:9}
\end{figure*}

\begin{figure}[h!]
\centering
\includegraphics[width=0.35\textwidth]{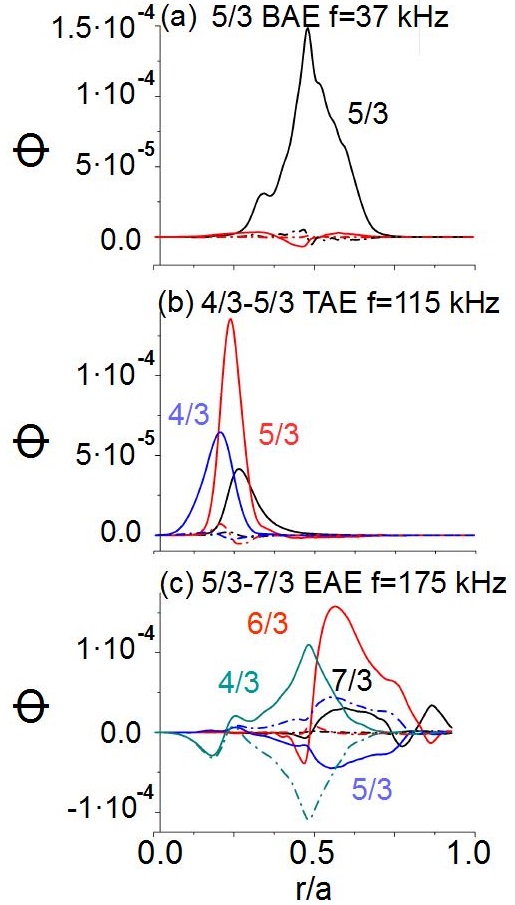}
\caption{Eigen-function of the $5/3$ BAE (a), $4/3-5/3$ TAE (b) and  $5/3-7/3$ EAE (c). Solid lines indicate the real component and the dot-dashed lines the imaginary component of the eigenfunction.}\label{FIG:10}
\end{figure}

Figure~\ref{FIG:9} shows the growth rate and frequency of the $n=1$ to $4$ CGDM, BAE, TAE and EAE for different $\beta_{f}$ values. The $n=1$ to $n=4$ CGDM are weakly affected by the EP destabilizing effect (panels a and e). The $n=2$ to $4$ BAEs are destabilized if the EP $\beta$ is $0.025$  (panels b and f), the $n=2$ TAE if the EP $\beta = 0.025$, the $n=3$ if the EP $\beta = 0.05$, the $n=4$ if the EP $\beta = 0.01$ (panels c and g), the $n=1$ EAE if the EP $\beta = 0.05$ as well as the $n=2$ and $n=3$ EAE if the EP $\beta = 0.025$ (panels d and h). Figure~\ref{FIG:10} shows the eigenfunction of the $n=3$ instabilities if $\beta_{f} = 0.05$. A $5/3$ BAE with a frequency of $37$ kHz is unstable in the middle plasma region (panel a), a $4/3-5/3$ TAE with a frequency of $115$ kHz in the inner plasma region (panel b) and a  $5/3 - 7/3$ EAE with a frequency of $175$ kHz in the middle plasma region (panel c).

\begin{figure}[h!]
\centering
\includegraphics[width=0.5\textwidth]{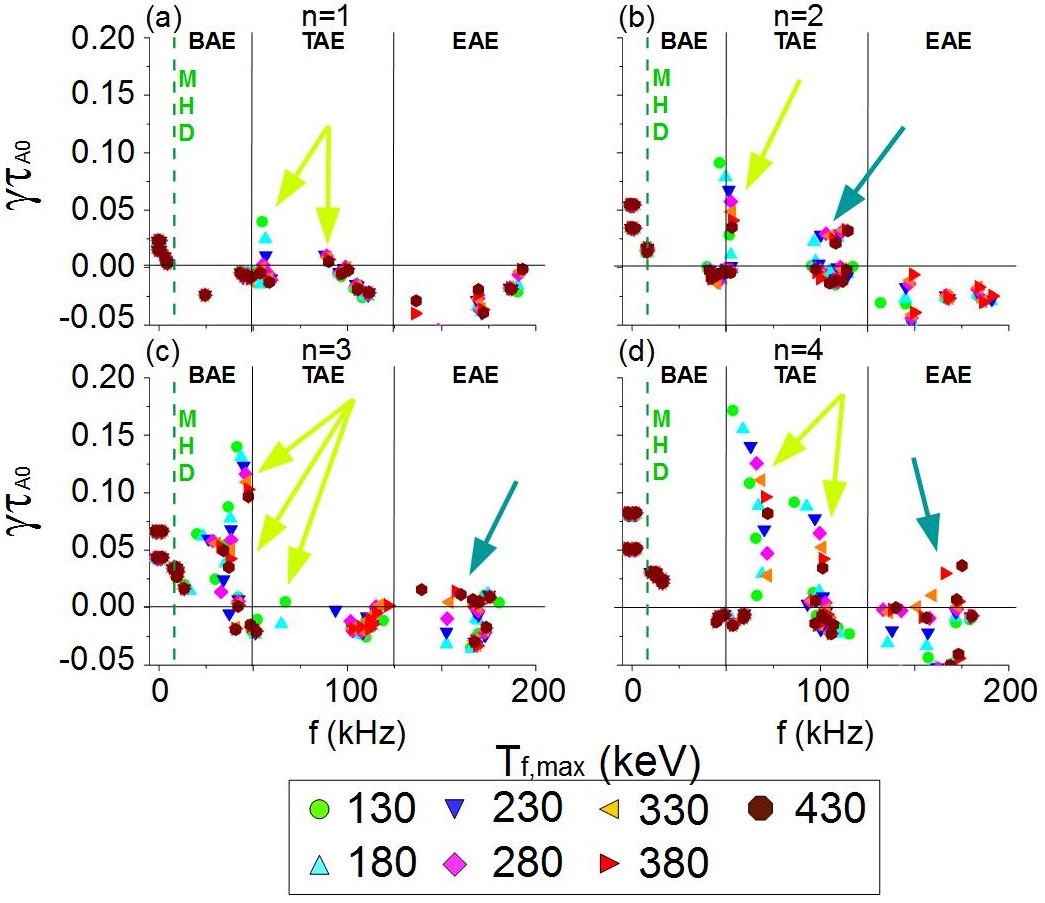}
\caption{Growth rate and frequency of the AEs for different EP energies if $\beta_{f} = 0.05$. The horizontal solid black line indicates the transition between stable (negative growth rate) and unstable (positive growth rate) modes. The vertical dashed green line indicates the range of frequencies of the AE (right side) and the pressure gradient driven modes (left side). The vertical solid black lines indicate the range of frequencies of the BAE, TAE and EAE. The yellow arrows indicates the modes that shows an increase of the growth rate as the EP energy decreases. The cyan arrows indicates the modes that show a decrease of  the growth rate as the EP energy decreases.}\label{FIG:11}
\end{figure}

Figure~\ref{FIG:11} shows the growth rate and frequency of the AEs for different EP energies if $\beta_{f} = 0.05$. The AE stability trends are the same compared to the simulations with an EP $\beta = 0.01$ (see fig.~\ref{FIG:5}) although the growth rate of the modes is higher. The unstable BAE and low frequency TAE (below $100$ kHz) if the EP energy decreases  (yellow arrows). On the other hand, the $n=2$ TAE and the $n=3$ to $n=4$ EAE show a decrease of the growth rate if the EP energy decreases (cyan arrows).

In short, the stability of the $n=1$ to $4$ CGDM is mainly unaffected by the EP. Also, the $\beta_{f}$ threshold is identified for several AEs, $0.025$ for the $n=2$ to $4$ BAEs, $n=2$ TAE and $n=2$ to $3$ EAEs, increasing to $0.05$ for the $n=3$ TAE and $n=1$ EAE.

\subsection{Effect of the on axis peaked EP profile on the AE stability}

In this section the stability of the AEs is analyzed for an on axis peaked EP profile. Consequently, the local maximum of the EP density and energy in the model is located at the magnetic axis (see fig.~\ref{FIG:1}e and f, black lines).

\begin{figure*}[h!]
\centering
\includegraphics[width=0.75\textwidth]{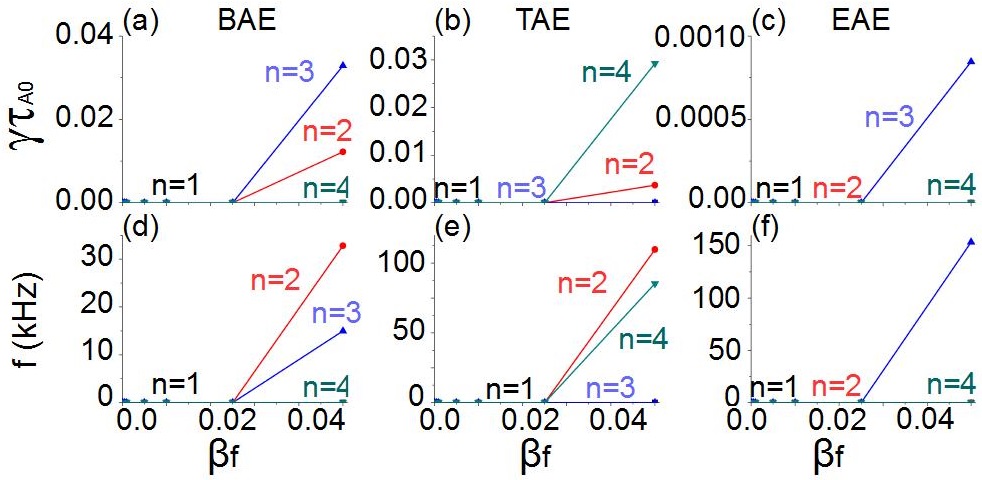}
\caption{Growth rate and frequency of the BAE (a and d), TAE (b and e) and EAE (c and f) for different $\beta_{f}$.}\label{FIG:12}
\end{figure*}

\begin{figure}[h!]
\centering
\includegraphics[width=0.45\textwidth]{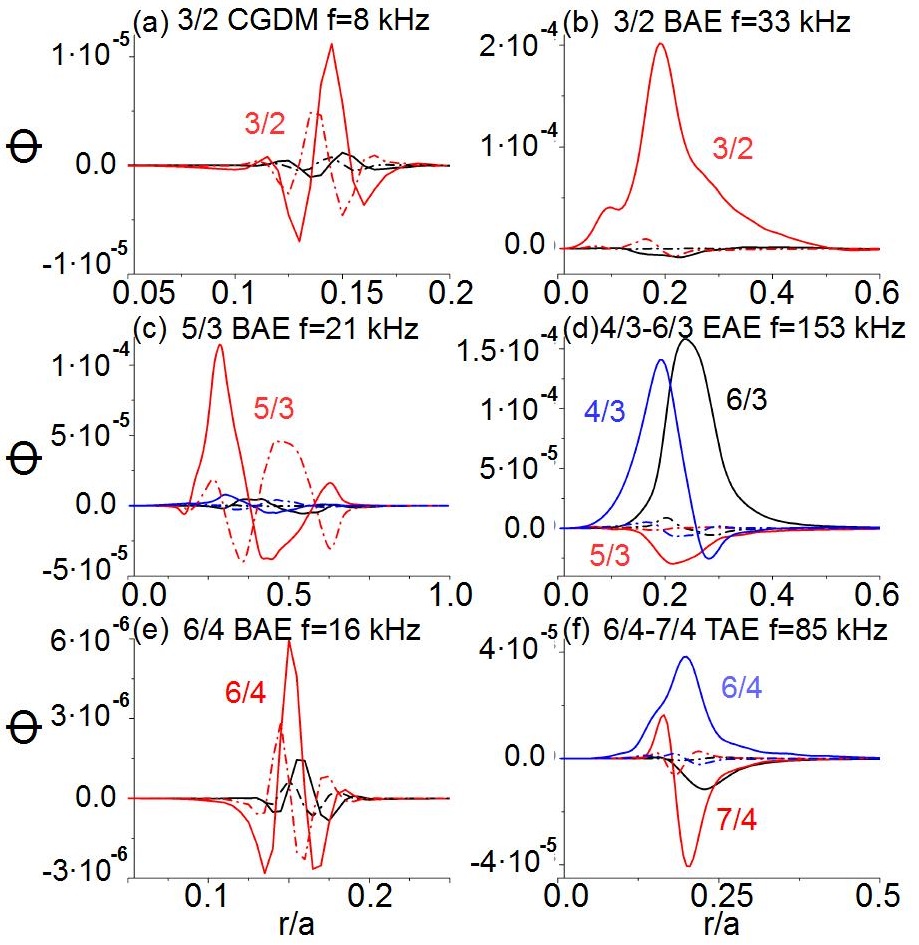}
\caption{Eigen-function of the $3/2$ CGDM (a), $3/2$ BAE (b), $5/3$ BAE (c), $4/3-6/3$ EAE (d), $6/4$ BAE (e) and $6/4-7/4$ TAE (f). Solid lines indicate the real component and the dot-dashed lines the imaginary component of the eigenfunction.}\label{FIG:13}
\end{figure}

Figure~\ref{FIG:12} shows the growth rate and frequency of the $n=1$ to $4$ BAE, TAE and EAE for different $\beta_{f}$ (the EP energy is fixed to $T_{f,max} = 280$ keV and the maxima of the EP density profile is modified). The $n=1$ BAE, TAE and EAE are stable up to $\beta_{f} = 0.05$. The $n=2$ and $3$ BAE are unstable if $\beta_{f} = 0.05$ in the range of frequencies from $15$ to $35$ kHz. The $n=2$ and $4$ TAE are unstable if $\beta_{f} = 0.05$ in the range of frequencies from $85$ to $115$ kHz. The $n=3$ EAE is marginal unstable if $\beta_{f} = 0.05$ with a frequency of $155$ kHz. It should be noted that, compared with the off-axis case, the AE stability is improved because the $\beta_{f}$ threshold is higher and the growth rate of the AEs is lower. The analysis of the RBM and CGDM stability is not shown because the growth rate is very similar to the off-axis case. Figure~\ref{FIG:13} shows the eigenfunction of the instabilities in the on-axis simulations if $\beta_{f} = 0.05$. The eigenfunction of the $n=2$ CGDM (panel a) is very similar to the off-axis simulations, the $3/2$ BAE with $33$ kHz is unstable in the inner plasma (panel b), the $5/3$ BAE with $21$ kHz in the middle plasma (panel c), the $4/3-6/3$ EAE with $153$ kHz in the inner plasma (panel d), the $6/4$ BAE with $16$ kHz in the inner plasma (panel e) and the $6/4-7/4$ TAE with $85$ kHz also in the inner plasma region (panel f).

Summarizing, the AE stability improves if the EP profile is peaked on axis with respect to the off axis case, because the $\beta_{f}$ threshold is higher and the growth rate of the AEs is lower. Also, if these AEs can coexist in the non-linear phase, the overlapping of the $n=2$ to $4$ BAEs in the inner plasma region can enhance the EP transport, although less with respect to the off-axis case.

\subsection{Stabilizing effect of the EP on the RBM}

\begin{figure}[h!]
\centering
\includegraphics[width=0.45\textwidth]{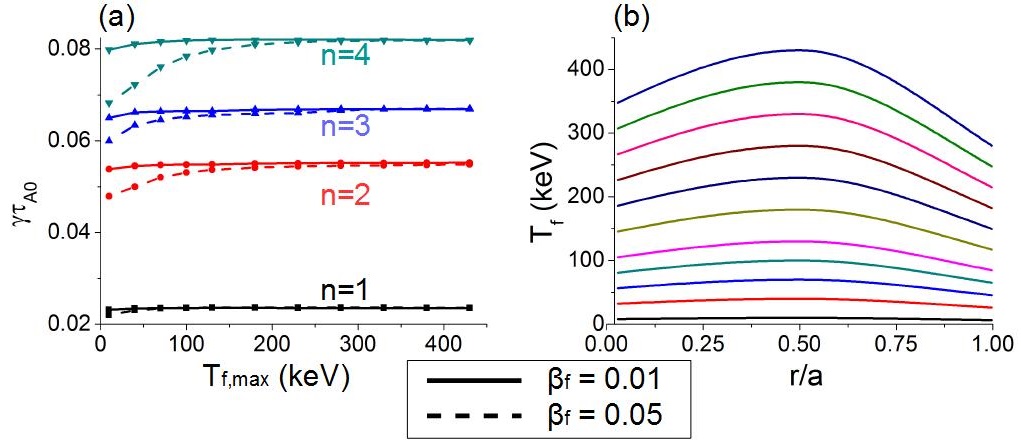}
\caption{(a) Growth rate of the RBM for different $\beta_{f}$ in the on-axis and off-axis cases. (b) Eigen-function of the $n=4$ RBM in the off-axis case if $\beta_{f} = 0.05$.}\label{FIG:14}
\end{figure}

The effect of the EP on the RBM is analyzed, studying how the RBM growth rate changes if the EP energy is modified for an EP $\beta$ of $0.01$ and $0.05$. The EP $\beta$ is fixed in the simulations as the EP energy changes by increasing/decreasing the EP density. Figure~\ref{FIG:14}a indicates the $n=1$ to $4$ RBM growth rate in the ITER-like inductive scenario (EP $\beta = 0.01$) and a configuration where the EP $\beta$ is increased up to $0.05$, scanning the EP energy from $10$ keV to $430$ keV ($S=5 \cdot 10^{6}$). Figure~\ref{FIG:14}b shows the different profiles of the EP energy used in the study (the ratio of the profile minimum and maximum is fixed). The growth rate of the RBM is weakly affected if the EP $\beta$ is $0.01$, showing only a small decrease if the EP energy is lower than $100$ keV. On the other hand, if the EP $\beta$ is $0.05$, the growth rate of the RBM meaningfully decrease for EP energies below $250$ keV, leading to a growth rate decrease up to a $15 \%$ in the simulations with an EP energy of $10$ keV. Consequently, the EP have a stabilizing effect on the RBM, enhanced if the population of EP with low energy increases. It should be noted that, because the EP $\beta$ is fixed in the simulations, a decrease of the EP energy is compensated by a proportional increase of the EP density and vice versa. A low energy EP results in a smaller $v_{th,f}/v_{A}$ ratio, leading to a resonance between the EP and the bulk plasma that improves the stability of the pressure gradient driven modes, that is to say, the EP population modifies the stability properties of the thermal plasma. It should be noted that the code TRANSP \cite{72} shows a local maxima of the EP energy at the pedestal of DIII-D plasma. On the other hand, below energies of $80$ keV, the distribution function of co-passing EP generated by the N-NBI show a limited population of EP in JT60U plasma, thus the required population of low energy EP to stabilize the RBM must be verified in JT60SA plasma. It should be noted that the EP generated by the P-NBI in JT-60SA plasma are in a range of energy below $50$ keV, although the population of passing/trapped EP should be verified, thus the stabilizing effect of the P-NBI must be also analyzed in detail for each EP population. Consequently, this kind of stabilizing resonance between EP and RBM could exist in JT-60SA plasma although must be confirmed.

In brief, there is a stabilizing effect of the EP on the RBM although it is caused only by EP at the end of the thermalization process. The existence of such a low energy EP population should be verified experimentally because co-passing EP generated by the N-NBI in JT60U plasma with energies below $80$ keV is expected to be relatively low.

\section{Hypothetical scenario with $q_{0}<1$ \label{sec:CaseB}}

In this section the stability of the CGDM and fish-bones is analyzed in a hypothetical JT-60SA scenario where the rational surfaces $q=1$ is resonant in the inner plasma region (see fig~\ref{FIG:2}).

\begin{figure}[h!]
\centering
\includegraphics[width=0.45\textwidth]{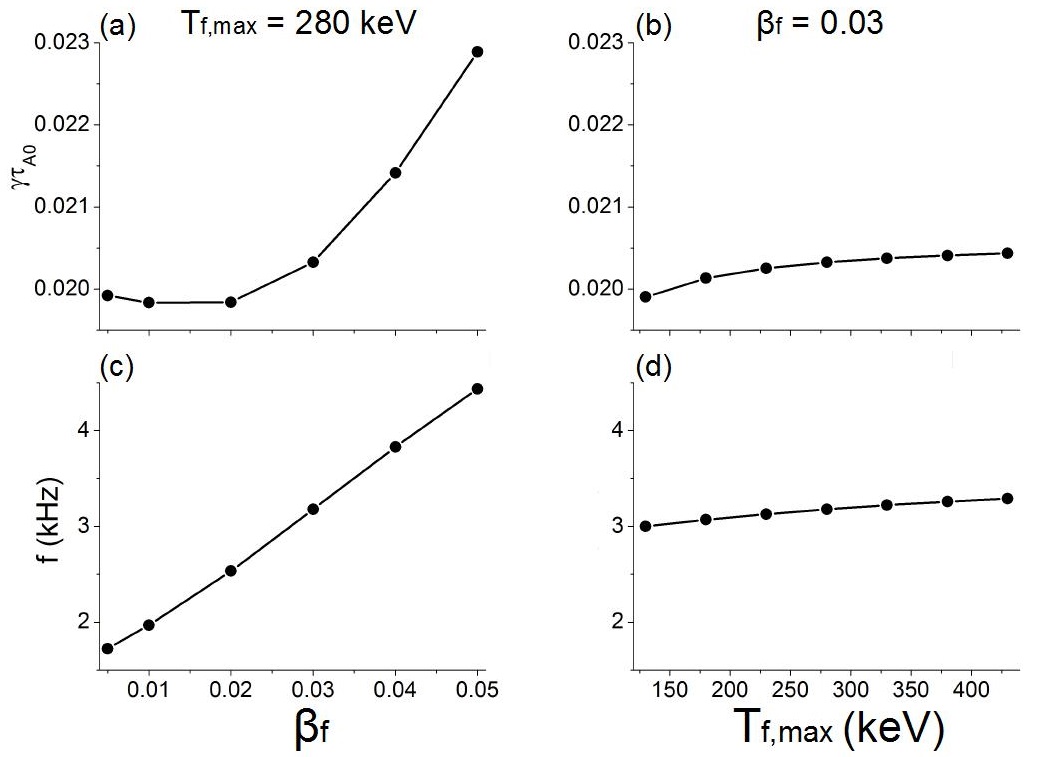}
\caption{(a) Growth rate and frequency of the $n=1$ mode for different $\beta_{f}$ (a and c) and EP energies (b and d) in the off-axis case.}\label{FIG:15}
\end{figure}

\begin{figure}[h!]
\centering
\includegraphics[width=0.45\textwidth]{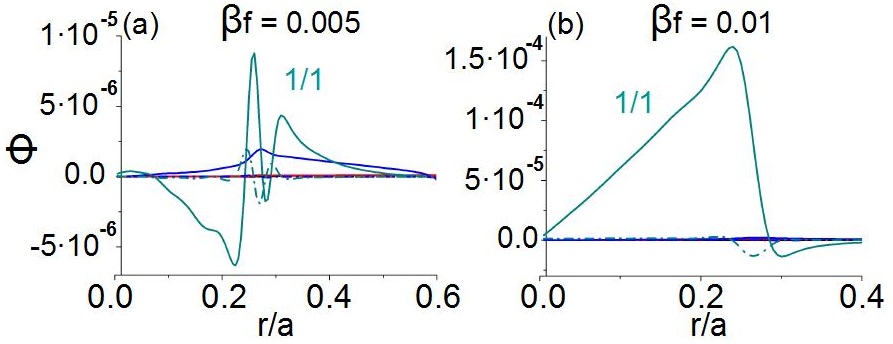}
\caption{Eigen-function of the $n=1$ mode if (a) $\beta_{f}=0.005$ and (b) $\beta_{f}=0.01$. Solid lines indicate the real component and the dot-dashed lines the imaginary component of the eigenfunction.}\label{FIG:16}
\end{figure}

Figure~\ref{FIG:15} shows the growth rate and frequency of the $n=1$ mode for different $\beta_{f}$ and EP energies. The analysis of the EP energy on the $n=1$ mode stability is performed scaling the maxima of the EP energy profile (see fig.~\ref{FIG:2}f). The growth rate and frequency of the $n=1$ mode increases with the $\beta_{f}$ above $0.02$, showing a different $\beta_{f}$ dependence compared to the $n=1$ CGDM (fig.~\ref{FIG:9}, panels a and c). On the other hand, the growth rate and frequency of the $n=1$ mode is weakly affected by the EP energy (panels b and d), showing a smaller increase of the growth rate as the EP energy increases compared to the EP $\beta$ study. Figure~\ref{FIG:16} shows the eigen-function of the $n=1$ mode for different $\beta_{f}$, indicating that for a $\beta_{f}=0.005$ the instability eigen-function is similar to the CGDM observed in the previous simulations, although if $\beta_{f}=0.01$ there is a transition to a different type of instability, showing a broader eigen-function between the magnetic axis and $r/a = 0.23$, where the $q=1$ rational surface is located. Consequently, this instability has several characteristics in common with the fish-bones observed in other Tokamak as DIII-D \cite{77} or JET \cite{78}, destabilized by passing/barely trapped EP \cite{79,81,81}. It should be noted that the fish-bones are also destabilized by trapped EP \cite{82} as it was observed in ASDEX \cite{83}. The analysis of the fish-bones stability in JT60SA plasma driven by trapped EP will be the topic of a future study.

In summary, JT-60SA operation scenarios where the rational surfaces $q=1$ is resonant in the inner plasma region, fish-bones could be triggered by passing/barely trapped EP.

\section{Conclusions and discussion \label{sec:conclusions}}

A set of linear simulations are performed by the FAR3d code to study the effects of the passing energetic particles on the stability of the RBM, CGDM, AEs and fish-bones in the ITER-like inductive scenario and hypothetical scenarios of JT-60SA with a resonant $q=1$ in the inner plasma region, which could lead to the prediction of the stability in N-NBI heated plasmas.

The simulations show the destabilization of $n=1$ to $4$ RBM at the plasma pedestal. The simulations performed using a magnetic Lundquist number similar to the experimental conditions ($S=7\cdot 10^7$) show stable $n=1$ to $4$ RBM at the plasma pedestal. The present study is performed using a fixed boundary model so the code results must be also confirmed by resistive free boundary codes. The study of the RBM stability in JT-60SA, particularly for scenarios that anticipate ITER and DEMO operation regimes, is a key issue to avoid the destabilization of type III ELMs at the plasma pedestal reducing the thermal fluxes to the plasma facing components. It should be noted that the most dangerous ELMs regarding the integrity of the device are the type I ELMs, triggered by unstable peeling ballooning modes (PBM) at the plasma pedestal. Previous numerical studies indicated that only high n PBM are unstable in ITER-like inductive scenarios \cite{3,73}. Consequently, the analysis of the PBM stability is out of the scope of the present study which is limited to low n modes.

The $n=4$ BAE, $n=2$ TAE and $n=4$ TAE could be destabilized in the ITER-like inductive scenario. Also, $n=1$ to $n=3$ BAEs can be triggered by EP with $T_{f,max} < 120$ keV, although the large population of low energy EP required to destabilized these AEs is not expected to be generated by the N-NBI, so these modes should be stable. A further analysis of the AE stability indicates that different AEs are destabilized above a given $\beta_{f}$ threshold, as it is shown in Table~\ref{Table:3}. If these modes coexist in the non linear (saturation) phase, the overlapping between instabilities located mainly between the inner and middle plasma region, can increase the EP transport and limit the JT-60SA performance due to an inefficient plasma heating. JT-60SA operation regimes with unstable AE/EPM must be avoided, because these instabilities deteriorate the device performance reducing the heating efficiency. The simulations that predict the destabilizing effect of the EP in JT-60SA operational scenarios is critical to identify the optimal JT-60SA magnetic field configuration, thermal plasma properties and NBI operational regime to improve the AE/EPM stability.

If the EP density and energy profiles are displaced to the magnetic axis in the ITER-like inductive scenario (on-axis case), the stability properties for the RBM and CGDM are not modified. On the other hand, the AE growth rate is lower and the $\beta_{f}$ destabilization threshold is higher (see table~\ref{Table:3}). The AEs are destabilized in the inner plasma region, showing a smaller overlapping between modes compared to the off-axis case. Consequently, the AE stability is improved in the on-axis case.

\begin{table}[t]
\centering
\begin{tabular}{c c c c}
\hline
Mode & Inst. & $\beta_{f}$ (Off-axis) & $\beta_{f}$ (On-axis) \\ \hline
$n=1$ & BAE & stable & stable \\
$n=2$ & BAE & $0.025$ & $0.05$ \\
$n=3$ & BAE & $0.025$ & $0.05$ \\
$n=4$ & BAE & $0.025$ & stable \\ 
$n=1$ & TAE & $0.05$ & stable \\
$n=2$ & TAE & $0.025$ & $0.05$ \\
$n=3$ & TAE & $0.05$ & stable \\
$n=4$ & TAE & $0.01$ & $0.05$ \\ 
$n=1$ & EAE & $0.05$ & stable \\
$n=2$ & EAE & $0.025$ & stable \\
$n=3$ & EAE & $0.025$ & $0.05$ \\
$n=4$ & EAE & stable & stable \\ 

\hline
\end{tabular}
\caption{EP $\beta$ threshold in the ITER-like inductive scenario and the on-axis case for the different AE families (fixed $T_{f,max} = 280$ keV).} \label{Table:3}
\end{table}

The simulations show an increase of the growth rate of the BAE and lower frequency TAE if the EP energy decreases, because the ratio $v_{th,f}/v_{A}$ is smaller as the $v_{th,f} \propto \sqrt{T_{f}}$ decreases, so the thermalized velocity of the EP is far from the super-Alfvenic condition $v_{th,f}/v_{A} > 1$. On the other hand, the simulations also show that the growth rate of the higher frequency TAE and EAE increase with the EP energy, pointing out an enhancement of the resonance if the ratio $v_{th,f}/v_{A}$ is closer to the super-Alfvenic condition. 

The simulations of the hypothetical JT-60SA scenario with a magnetic configuration where the safety factor is lower than the unit at the magnetic axis, show the destabilization of an $n=1$ EPM similar to fish-bones. The growth rate and frequency of the $n=1$ EPM increases with $\beta_{f}$, a feature not observed in the CGDM calculated by the code. Also, the $n=1$ mode shows a transition between different instabilities if $\beta_{f}=0.01$, where the instability eigen-function changes: the eigen-function width increases, located between the magnetic axis and $r/a=0.23$ where the $q=1$ rational surface resonates in the inner plasma. The transition between $n=1$ CGDM to fish-bone like $n=1$ EPM  is similar to the transition observed in DIII-D plasma with respect to the safety factor value at the magnetic axis \cite{84}.

The simulations show a stabilizing effect of the EP on the RBM. The stabilization effect is enhanced if the population of EP with low energies (less than $250$ keV if the EP $\beta = 0.05$) at the plasma pedestal increases. It should be noted that the EPs driven by the P-NBI are generated in a range of energies that could also have a stabilizing effect on the RBM. This possibility will be explored in future studies. In principle, this stabilizing kinetic effect on the RBM should be verified in other tokamaks as DIII-D or ASDEX, first confirming if such low energy EP population at the plasma pedestal exists.

The drive of the AE modes is determined by the gradient of the phase space distribution towards the direction which particles can move at the resonant condition. The gradient depends on the phase space shape of the distribution function. Thus, simulations using an anisotropic slowing down distribution function is required to confirm whether the instabilities calculated could be destabilized in a JT-60SA plasma. This is future work.

Future studies dedicated to analyze the stability of JT-60SA plasma will include different JT-60SA operation scenarios, such as non inductive or steady state scenarios, adding in the analysis not only the destabilizing effect of the passing EP, also the effect of the trapped EP.

\section*{Appendix}

\subsection*{EP distribution function}

The EP distribution in the simulations is an Maxwellian which has the same second moment, the effective EP temperature, with the slowing down distribution as defined below:
\begin{eqnarray} 
\label{eq:SD}
f_{SD} = \frac{\tau_{s}}{v^{3} + v^{3}_{c}} \left( \frac{v^{3}}{v^{3} + v^{3}_{c}}\right)^{b/3} \nonumber\\
\hspace{2.5cm} \int^{v}_{\infty} \left(\frac{v'^{3}}{v'^{3} + v^{3}_{c}}\right)^{-b/3} S (v') dv'^{3}
\end{eqnarray} 
and the Maxwellian distribution as:
\begin{equation}
\label{eq:Max}
f_{Max} = N_{M}e^{\frac{-mv^{2}}{K_{B}T}}
\end{equation}
where $v_{c} = (3\sqrt{\pi} m_{e}/4m_{i})^{1/3} \cdot v_{e}$, with $m_{e}$ the electron mass, $m_{i}$ the ion mass and $v_{e}$ the electron velocity, and $v_{EP,NBI} = \sqrt{2E_{EP,NBI}/m_{EP,NBI}}$ the beam particles velocity with $E_{EP,NBI}$ the beam particles energy and $m_{EP,NBI}$ the beam particles mass. $\tau_{s}$ is the slowing down time and $b$ is a dimensionless parameter that indicates the effect of transport. For simplicity, we consider the effect of the transport negligible, a mono-energetic source and a isotropic distribution function, thus:
\begin{equation}
f_{SD} = \frac{S_{0} \tau_{s}}{4\pi} \frac{1}{v^{3} + v^{3}_{c}} H(v - v_{EP,NBI})
\end{equation}
The consequence of these simplifications is that the EP model cannot reproduce the destabilization caused by highly anisotropic beams or ICRF driven EP, although the destabilizing effect of passing particles generated by a tangential NBI  is reproduced.
The averaged square velocity of the slowing down and Maxwellian distribution is selected to be the same ($\langle v^{2} \rangle_{Max} = \langle v^{2} \rangle_{SD}$), where the averaged square velocity is defined as:
\begin{equation}
\langle v^{2} \rangle = \frac{\int fv^{2}dv^{3}}{\int fdv^{3}}
\end{equation}
The assumption of the model is that the averaged Maxwellian is similar to the thermalized velocity of the EP, thus:
\begin{eqnarray} 
\langle v^{2} \rangle_{Max} = \frac{\left(\frac{K_{B}T_{f}}{m_{f}}\right)^{5/2} \int^{\infty}_{0} e^{-x^{2}}x^{5/2}dx}{\left(\frac{K_{B}T_{f}}{m_{f}}\right)^{3/2} \int^{\infty}_{0} e^{-x^{2}}x^{3/2}dx} \nonumber\\
\hspace{1.5cm} \approx \frac{K_{B}T_{f}}{m_{f}} \approx v_{th,f}^{2} 
\end{eqnarray} 
with $x^{2} = \frac{m_{f} v^{2}}{K_{B}T_{f}}$, where:
\begin{equation}
\langle v^{2} \rangle_{SD} = \frac{\int^{v_{EP,NBI}}_{0} \frac{v^{4}dv}{v^{3} + v^{3}_{c}}} {{\int^{v_{EP,NBI}}_{0} \frac{v^{2}dv}{v^{3} + v^{3}_{c}}}} = v^2_{c} \frac{\int^{v_{EP,NBI}/v_{c}}_{0} \frac{x^{4}dx}{x^{3} + 1}} {{\int^{v_{EP,NBI}/v_{c}}_{0} \frac{x^{2}dx}{x^{3} + 1}}}
\end{equation}
with $x = v / v_{c}$. Consequently, if the electron temperature is $1$ keV:
\begin{equation}
T_{f} = 0.573 E_{NBI}
\end{equation}

\ack

The authors want to thank QST Naka technical staff for their contributions in the study of JT-60SA operation scenarios. The authors also wish to acknowledge N. Aiba for fruitful discussion. This work is supported in part by NIFS under contract NIFS07KLPH004.

\hfill \break


\begin{thebibliography}{10}

\bibitem{1} S. Ishida et al {\it Nucl. Fusion}, {\bf 51}, 094018, (2011). 
\bibitem{2} JT-60SA Research Plan 2016 {\it www.jt60sa.org/pdfs/JT-60SA\_Res\_Plan.pdf}
\bibitem{3} G. Giruzzi et al {\it Nucl. Fusion}, {\bf 57}, 085001, (2017). 
\bibitem{4} M. Shimada et al {\it Progress in the ITER physics basis chapter 1. Overview and summary} {\it Nucl. Fusion}, {\bf 47}, S1, (2007). 
\bibitem{5} G. Federici et al {\it Fusion Eng. Des.}, {\bf 89}, 882, (2014). 
\bibitem{6} G. Federici et al {\it Nucl. Fusion}, {\bf 57}, 092002, (2017). 
\bibitem{7} Y. Kamada et al {\it Nucl. Fusion}, {\bf 51}, 073011, (2011).  
\bibitem{8} A. C. C. Sips et al {\it Phys. of Plasma}, {\bf 22}, 021804, (2015).  
\bibitem{9} L. Garzotti et al {\it Nucl. Fusion}, {\bf 58}, 026029, (2018).
\bibitem{10} F.M. Poli, et al {\it Nucl. Fusion}, {\bf 52}, 063027 (2012).
\bibitem{11} D.N. Hill et al {\it Nucl. Fusion}, {\bf 53}, 104001, (2013).   
\bibitem{12} S.H. Kim, et al, {\it Nucl. Fusion}, {\bf 56}, 126002 (2016).
\bibitem{13} T. Casper et al {\it Nucl. Fusion}, {\bf 54}, 013005, (2015).
\bibitem{14} M. Hanada et al {\it Fusion Eng. Des.}, {\bf 86}, 835, (2011). 
\bibitem{15} A. Kojima et al {\it Nucl. Fusion}, {\bf 55}, 063006, (2015).
\bibitem{16} K. Toi et al {\it Nucl. Fusion}, {\bf 44}, 217, (2004). 
\bibitem{17} Y. Samamoto et al {\it Nucl. Fusion}, {\bf 45}, 326, (2005). 
\bibitem{18} J. R. Wilson et al {\it Bull. Am. Phys. Soc.}, {\bf 37}, 1380, (1992).  
\bibitem{19} K. L. Wong et al {\it Phys. Rev. Lett.}, {\bf 66}, 1874, (1991). 
\bibitem{20} S. E. Sharapov et al {\it Nucl. Fusion}, {\bf 39}, 373, (1999). 
\bibitem{21} W. W. Heidbrink et al {\it Nucl. Fusion}, {\bf 31}, 1635, (1992).  
\bibitem{22} H. H. Duong et al {\it Nucl. Fusion}, {\bf 33}, 749, (1993).  
\bibitem{23} A. Bierwage et al {\it Plasma Phys. Control Fusion}, {\bf 59}, 125008, (2017).  
\bibitem{24} H. Kimura et al {\it Nucl. Fusion}, {\bf 38}, 1303, (1998).  
\bibitem{25} Y. Kusama et al {\it Nucl. Fusion}, {\bf 39}, 1837, (1999).  
\bibitem{26} K. Shinohara et al {\it Nucl. Fusion}, {\bf 41}, 603, (2001). 
\bibitem{27} K. Shinohara et al {\it Nucl. Fusion}, {\bf 42}, 942, (2002). 
\bibitem{28} M. Takechi et al {\it Phys. of Plasma}, {\bf 12}, 082509, (2005). 
\bibitem{29} D. A. D’Ippolito et al {\it Plasma Phys.}, {\bf 22}, 1091, (1980). 
\bibitem{30} B. Van der Holst et al {\it Phys. Rev. Lett.}, {\bf 84}, 2865, (2000).
\bibitem{31} C. Kieras et al {\it Plasma Phys.}, {\bf 28}, 395, (1982).  
\bibitem{32} C. Z. Cheng et al {\it Phys. Fluids}, {\bf 29}, 3695, (1986).  
\bibitem{33} Zhixuan Wang et al {\it Phys. Fluids}, {\bf 22}, 022509, (2015).
\bibitem{34} A. D. Turnbull et al {\it Phys. Fluids B}, {\bf 5}, 2546, (1993).
\bibitem{35} D. A. Spong et al {\it Nucl. Fusion}, {\bf 53}, 053008, (2013).  
\bibitem{36} T. E. Evans et al {\it Phys. Rev. Lett.}, {\bf 53}, 1743, (1984).
\bibitem{37} R. Betti et al  {\it Phys. Fluids B}, {\bf 4}, 1465, (1992).
\bibitem{38} G. J. Kramer et al {\it Phys. Rev. Lett.}, {\bf 80}, 2594, (1998). 
\bibitem{39} R. Betti et al {\it Phys. Fluids B}, {\bf 3}, 1865, (1991).
\bibitem{40} E. M. Bass et al {\it Phys. Plasmas}, {\bf 20}, 012508 (2013).
\bibitem{41} L. Chen et al {\it Phys. Plasmas}, {\bf 1}, 1519, (1994).
\bibitem{42} T. Ido et al {\it Nucl. Fusion}, {\bf 51}, 073046, (2011).
\bibitem{43} T. Ido et al {\it Nucl. Fusion}, {\bf 55}, 083024, (2015).
\bibitem{44} H. Wang et al {\it Phys. Plasmas}, {\bf 22}, 092507, (2015).
\bibitem{45} H. Wang et al {\it Phys. Rev. Lett.}, {\bf 120}, 175001, (2018).
\bibitem{46} K. Nagaoka et al {\it Nucl. Fusion}, {\bf 51}, 083022, (2011).   
\bibitem{47} X. D. Du et al {\it Phys. Rev. Lett.}, {\bf 114}, 155003, (2015). 
\bibitem{48} J. Varela, J. et al {\it Nucl. Fusion}, {\bf 59}, 046008, (2019).
\bibitem{49} M. F. F. Nave et al {\it Nucl. Fusion}, {\bf 31}, 697, (1991).    
\bibitem{50} J. W. Connor et al {\it Plasma Phys. Control. Fusion}, {\bf 27}, 1509, (1985).  
\bibitem{51} G. S. Bevli et al {\it Phys. Rev. Lett. E}, {\bf 48}, 1509, (1993). 
\bibitem{52} C. Bourdelle et al {\it Plasma Phys. Control Fusion}, {\bf 54}, 115003, (2012). 
\bibitem{53} J. W. Connor et al {\it Plasma Phys. Control. Fusion}, {\bf 40}, 191, (1998).
\bibitem{54} W. Suttrop {\it Plasma Phys. Control. Fusion}, {\bf 42}, A1, (2000). 
\bibitem{55} L. Garcia \textit{Proceedings of the 25th EPS International Conference, Prague, 1998}, VOL. 22A, Part II, p. 1757.
\bibitem{56} L. A. Charlton et al {\it Journal of Comp. Physics}, {\bf 63}, 107, (1986).
\bibitem{57} L. A. Charlton et al {\it Journal of Comp. Physics}, {\bf 86}, 270, (1990).
\bibitem{58} S. P. Hirshman et  al {\it Phys. Fluids}, {\bf 26}, 3553, (1983).
\bibitem{59} D. A. Spong et al {\it Phys. Fluids B}, {\bf 4}, 3316, (1992).  
\bibitem{60} C. L. Hedrick et al {\it Phys. Fluids B}, {\bf 4}, 3869, (1992).  
\bibitem{61} D. A. Spong et al {\it Nucl. Fusion}, {\bf 53}, 053008, (2013).  
\bibitem{62} L. Garcia et al {\it Phys. Fluids B}, {\bf 2}, 2162, (1990).
\bibitem{63} G. W. Hammett et  al {\it Phys. Rev. Lett.}, {\bf 64}, 3019, (1990).
\bibitem{64} F. Zonca et al {\it Plasma Phys. Control. Fusion}, {\bf 38}, 2011, (1996). 
\bibitem{65} W. Deng et al {\it Phys. Plasmas}, {\bf 17}, 112504, (2010). 
\bibitem{66} A.H. Boozer {\it Phys. Fluids}, {\bf 25}, 520, (1982).
\bibitem{67} J. Varela et al {\it Nucl. Fusion}, {\bf 58}, 076017, (2018). 
\bibitem{68} J. Varela et al {\it Nucl. Fusion}, {\bf 57}, 046018, (2017). 
\bibitem{69} J. Varela et al {\it Nucl. Fusion}, {\bf 57}, 126019, (2017). 
\bibitem{70} J. Varela et al {\it Nucl. Fusion}, {\bf 59}, 046017, (2019). 
\bibitem{71} J. Varela et al {\it Nucl. Fusion}, {\bf 59}, 076036, (2019). 
\bibitem{72} N. C. Logan et al {\it Fusion Sci. Technol}, {\bf 74}, 125, (2018).
\bibitem{73} N. Aiba et al {\it Plasma Phys. Control. Fusion}, {\bf 60}, 014032, (2018).
\bibitem{74} L. Pigatto et al {\it Nucl. Fusion}, {\bf 59}, 106028, (2019).
\bibitem{75} D. Spong et al {\it Phys. Plasmas}, {\bf 10}, 3217, (2003). 
\bibitem{76} M. S. Chu et al {\it Phys. Fluids}, {\bf B4}, 3713, (1992).
\bibitem{77} W.W. Heidbrink et al {\it Nucl. Fusion}, {\bf 30}, 1015, (1990).
\bibitem{78} V.G. Kiptily et al {\it Nucl. Fusion}, {\bf 58}, 014003, (2018).
\bibitem{79} B. Coppi et al {\it Phys. Rev. Lett.}, {\bf 57}, 2272, (1986).
\bibitem{80} R. Betti et al {\it Phys. Rev. Lett.}, {\bf 70}, 3428, (1993).
\bibitem{81} W.W. Heidbrink et al {\it Nucl. Fusion}, {\bf 34}, 535, (1994).
\bibitem{82} L. Chen et al {\it Phys. Rev. Lett.}, {\bf 52}, 1122, (1984).
\bibitem{83} T. Kass et al {\it Nucl. Fusion}, {\bf 38}, 807, (1998).
\bibitem{84} R. Zhen-Zhen et al {\it Phys. Plasmas}, {\bf 25}, 122504, (2018). 


\end{thebibliography}
\end{document}